\documentclass[A4paper,traditabstract]{aa}
\usepackage{graphicx}
\usepackage{txfonts}

\def\bege{\begin{equation}}
\def\ende{\end{equation}}
\def\gsim{\lower.4ex\hbox{$\;\buildrel >\over{\scriptstyle\sim}\;$}}
\def\lsim{\lower.4ex\hbox{$\;\buildrel <\over{\scriptstyle\sim}\;$}}

\begin{document}
   \title{The differential rotation of G dwarfs}

   \author{
          M.~K\"uker\inst{1} \and  G.~R\"udiger\inst{1} \and
        L.L.Kitchatinov\inst{2,3} 
           }

   \institute{Astrophysikalisches Institut Potsdam, 
   An der Sternwarte 16, D-14482 Potsdam\\
   \email{mkueker@aip.de, gruediger@aip.de}
         \and
         Institute of Solar-Terrestrial Physics, PO Box 291, 
         Irkutsk 664033, Russia \\
   \email{kit@iszf.irk.ru}
         \and
         Pulkovo Astronomical Observatory, St. Petersburg 196140, Russia
             }

   \date{\today }
\authorrunning{M. K\"uker, G. R\"udiger \& L.L. Kitchatinov}
 \titlerunning{Differential rotation of G stars}

\abstract{
A series of stellar models of spectral type G  is computed to  
study  the rotation laws resulting from  mean-field equations.
 The rotation laws of the slowly rotating  Sun, the fast rotating MOST stars $\epsilon$ Eri and $\kappa^1$ Cet and the rapid  rotators R58 and LQ Lup 
can easily be reproduced. We also find  that differences in the depth of the  convection zone cause large differences in the surface rotation law and that the extreme surface shear of 
 HD 171488 can only be explained  with a artificially  shallow convection layer.\\
We also check the thermal wind equilibrium in fast-rotating G dwarfs and find  that the polar subrotation (${\rm d} \Omega/{\rm d} z<0$) is due to the barocline effect and that the equatorial superrotation (${\rm d} \Omega/{\rm d} r>0$) is due to the $\Lambda$ effect as part of the  Reynolds stresses. In the bulk of the convection zones where the meridional flow is slow and smooth the thermal wind equilibrium actually holds between the centrifugal and the pressure forces. It does not hold, however,  in the bounding shear  layers including the equatorial region where the Reynolds stresses dominate.}  
\maketitle

\section{Introduction}
Many stars show signs of differential rotation. The solar equator rotates with a shorter period than the polar caps. The difference of 132 nHz between the rotation periods found by \cite{ulrich88} corresponds to a difference of 0.07 rad/day between the respective angular velocities or a lapping time of 88 days. Helioseismology has found that this pattern persists throughout the whole convection zone but not in the radiative zone below (\cite{thompson03}). Stellar differential rotation can be inferred from the light curves of rotating spotted stars (see, e.g.  Henry et al.~1995; Messina \& Guinan 2003), 
from monitoring magnetic activity (Donahue et al.~1996), 
spectroscopically (\cite{reiners03}), or by Doppler imaging (\cite{barnes2000}; \cite{donati2000}). 

While several studies   have found a systematic dependence of the surface differential rotation on the rotation rate, no such dependence is  found when the samples are combined (Hall 1991; \cite{barnes05}). Moreover, measurements of surface differential rotation of the rapidly rotating K dwarfs PZ Tel and AB Dor with the Doppler imaging technique show that stars rotating much faster than the Sun show very similar surface shear values  -- as first predicted by Kitchatinov \& R\"udiger (1999). 
PZ Tel is a young K dwarf with a rotation period of 0.95 days. Its surface differential rotation $\delta \Omega = 0.075$ rad/day is remarkably close to that of the Sun (\cite{barnes2000}). AB Dor is a rapidly rotating K0 dwarf with a rotation period of 0.51 days. Its surface differential rotation was found to vary with time between 0.09 and 0.05 rad/day (\cite{cameron02}).

Barnes et al. (2005) proposed a dependence on the effective temperature (albeit with large scatter) and found the power law
\bege  
   \delta \Omega = \Omega_{\rm eq}-\Omega_{\rm pole} \propto T_{\rm eff}^{8.92\pm0.31}
   \label{barneslaw}
\ende 
 where $\delta \Omega$ is the difference between the angular velocities at the equator and the polar caps \footnote{the strength of the differential rotation is also  expressed in terms of the lapping time between the equator and the poles,
$
 P_{\rm lap} = {2 \pi}/{\delta \Omega}
$}.
The power law (\ref{barneslaw}) was confirmed by the findings using  spectroscopic methods (Reiners 2006). The light curves of the stars $\epsilon$ Eri and $\kappa^1$ Cet recorded by the MOST satellite (\cite{croll06}; \cite{walker07}) both indicate with  $\delta\Omega\simeq 0.062$  and $\delta\Omega\simeq 0.064$ a very similar  
equator-pole difference of the rotation rate as the Sun.
Also the value of 0.11 for the young G star CoRoT-2a (\cite{froehlich09}) well fits the common picture of a rotation-independent surface shear for G stars. 

It   is challenged, however,  by recent observations of the young G dwarfs LQ Lup, R58 and HD171488.  
Marsden et al.~(2005) report a surface differential rotation of $0.025 \pm 0.015$ rad/day for R58 while \cite{jeffers09b} find the much larger value of $0.138 \pm 0.011$ rad/day. Donati et al. (2000) find a surface differential rotation $\delta \Omega = 0.12 \pm 0.02$ rad/day for LQ Lup.
\cite{jeffers09a} determined the surface rotation of HD 171488 using the Zeeman-Doppler imaging technique and found a very strong surface differential rotation of 0.5 rad/day.
Marsden et al. (2006) report a smaller but still large value of $0.402\pm0.044$ rad/day. Huber et al.~(2009) found no evidence for differential rotation at all but could not rule it out either. Jeffers et al.~(2010) confirmed the findings of Marsden et al.~(2006).

The values found for LQ Lup and R58 are in line with the Barnes et al. picture but the large values found for  HD 171488 are not. The three G dwarfs are similar in their ages, effective temperatures and radii yet HD 171488 shows a much stronger differential rotation than the other two stars.
The studies mentioned have focused on rotation rate and effective temperature as the properties determining the surface differential rotation. As the stars are very similar by their stellar parameters such as age and effective temperature, we ask if there could be a difference in their internal structure that would cause the observed difference in their surface rotation. All three stars have just reached the zero age main sequence or are approaching it. Given the rapid retreat of the convection zone in the final part of the pre-main sequence evolution and the uncertainty of the stellar age we ask if the strong differential rotation  of HD 1714888 can be explained by a different depth of its outer convection zone.
   
In the following we compute model convection zones of different depth and their large-scale gas motions, i.e.~rotation and meridional flow.
The models are based on the mean field formulation of fluid dynamics which has been very successful for the Sun, where the models reproduce the surface rotation, the internal in the convection zone, and the surface meridional flow very well (Kitchatinov \& R\"udiger (1999, KR99), K\"uker \& Stix (2001, KS01)). Models have been constructed for a variety of stars with spectral types from M to F (K\"uker \& R\"udiger 2008).

A new scheme allows the computation of stellar rotation laws and meridional flow patterns based on a mean-field model of the large-scale flows in stellar convection zones also for fast  rotation rates when narrow boundary layers exist. It assumes strict spherical symmetry for the basic stratification, ignoring any flattening that might occur for very fast rotation. However, the impact of rotation on the thermal structure can be taken into account by including a gravity darkening term in the heat transport equation so that the model remains applicable even for moderately flattened stars.
\section{\label{theory}Theory of stellar rotation laws}
While thermal convection is driven by the stratification of the star, the convective gas motions carry angular momentum as well as heat. Momentum transport by turbulent motions is known as Reynolds stress and can be described as a turbulent viscosity in  case of the simplest  shear flows. In a rotating, stratified convection zone the Reynolds stress is anisotropic and its azimuthal components have a contribution proportional to the angular velocity, $\Omega$, itself rather than its gradient. This form of Reynolds stress (with `$\Lambda$ effect') is not compatible with solid body rotation. 

\subsection{Basic equations}
Our model consists of a 1D background model which assumes hydrostatic equilibrium and spherical symmetry and a system of partial differential equations that describe the convective heat flux, the transport of angular momentum, and the meridional flow, assuming axisymmetry. 
The equation for the convective heat transport then reads
\begin{equation} \label{heat}
    \nabla \cdot  (\vec{F}^{\rm conv} + \vec{F}^{\rm rad}) - \rho T \vec{\bar{u}} \cdot \nabla{S} = 0,
\end{equation}
where $\vec{F}^{\rm conv}$ and $\vec{F}^{\rm rad}$ are the convective and radiative heat flux, respectively,
$\rho$ the mass density,  $\vec{\bar{u}}$ the mean gas velocity, $T$ the gas temperature. $S$ is the specific entropy,
 \begin{equation}
  S=C_v \log\left(\frac{p}{\rho^\gamma}\right),
\end{equation}
where $p$ is the gas pressure and $C_v$ the specific heat at constant volume and $\gamma$ the adiabate index.
In  spherical polar coordinates the Reynolds equation can be rewritten as a system of two partial differential equations. The azimuthal component of the Reynolds equation expresses the conservation of angular momentum,
\begin{equation} \label{omega}
   \nabla \cdot \vec{t} = 0,
\end{equation}
where $\vec{t}$ is the angular momentum flux vector 
\begin{equation}
t_i =  r \sin \theta (\rho r \sin \theta \Omega {u}_i^{\rm m}
    + \rho Q_{i \phi}),
\end{equation}
with the two transporters of angular momentum i) the meridional flow  ($u^{\rm m}$) and ii) the zonal flux of the turbulent angular momentum  ($Q_{i\phi}$).

The equation for the meridional flow is derived from the Reynolds equation by taking the azimuthal component of its curl:
\begin{equation}
 \label{curl} 
  \left[ \nabla \times 
	   \frac{1}{\rho}\nabla \cdot R \right ]_\phi 
	   + r \sin \theta \frac{\partial \Omega^2}{\partial z} 
	  + \frac{1}{\rho^2}(\nabla \rho \times \nabla p)_\phi + \dots =0
\end{equation}
where $R_{ij}=-\rho Q_{ij}$ is the Reynolds stress and ${\partial}/{\partial z}=\cos \theta {\partial}/{\partial r}-\sin \theta {\partial}/{\partial \theta}$ is the derivative along the axis of rotation. For very fast rotation the second term on the LHS of Eq.~(\ref{curl}) dominates and the rotation rate is constant along cylindrical surfaces, as stated by the Taylor-Proudman theorem.
\subsection{Transport coefficients}
 If spherical polar coordinates are chosen, the $\Lambda$ effect appears in two components of the Reynolds stress:
\begin{eqnarray}
 Q_{r \phi} &=& Q^{\rm visc}_{r \phi} + V \nu_{\rm t} \sin \theta \Omega \\
 Q_{\theta \phi} &=& Q^{\rm visc}_{\theta \phi} + H \nu_{\rm t} \cos \theta \Omega
\end{eqnarray}  
Here, $Q^{\rm visc}_{r \phi}$ and $Q^{\rm visc}_{\theta \phi}$ contain only first order derivatives of $\Omega$ with respect to $r$ and $\theta$ and therefore vanish for uniform rotation. The coefficients $V$ and $H$ refer to the vertical and horizontal part of the $\Lambda$ effect while $\nu_{\rm t}$ represents the eddy viscosity in the convection zone. As the corresponding terms contain the angular velocity $\Omega$ itself, they do not vanish for rigid rotation. Thus, rigid rotation is not stress-free.

For slow rotation, the convective heat flux is proportional to the gradient of the specific entropy:
\bege \label{heat_iso}
   F^{\rm conv}_i= - \rho T \chi_{\rm t} \frac{\partial S}{\partial x_i},
\ende
where  $\chi_{\rm t}$ the turbulent heat conductivity. The turbulent heat conductivity is determined by the convection velocity $u_c$ and the convective turnover time $\tau_{\rm c}$:
\bege
   \chi_{\rm t} = \frac{1}{3} \tau_{\rm c} u_{\rm c}^2
\ende   
Using standard mixing-length theory we find
\bege
  u_{\rm c}^2 = -\frac{l_{\rm m}^2 g}{4 C_p} \frac{\partial S}{\partial r}
\ende
and 
\bege
 \tau_{\rm c} = \frac{l_{\rm m}}{u_{\rm c}}
\ende   
where $l_{\rm m}$ is the mixing length.
Equation~(\ref{heat_iso}) describes a strictly radial, diffusive heat flux. 
In a rotating convection zone the heat flux is not strictly aligned with the entropy gradient, i.e.
\bege
  F^{\rm conv}_i = -\rho T \chi_t \Phi_{ij} \frac{\partial S}{\partial x_j}
\ende
(R\"udiger 1989, \cite{kit94}), where the dimensionless coefficients $\Phi_{ij}$ are functions of the Coriolis number $\Omega^*=2 \tau_c \Omega$ fulfilling $|\Phi_{ij}|\le1$. The flux vector is then tilted towards the rotation axis.

The radiative heat flux is given by
 \begin{equation}
 F_i^{\rm rad} = - \frac{16 \sigma T^3\nabla_i T}{3 \kappa \rho}, 
 \end{equation}
where $\sigma$ is the Stefan-Boltzmann constant and $\kappa$ the opacity. 

\subsection{Background model}
We assume that the star is essentially in hydrostatic equilibrium and that all gas motions constitute small perturbations of that equilibrium which do not change the structure of the star. We particularly assume that no net gas transport occurs:
\bege
\nabla \cdot (\rho \vec{\bar{u}}) = 0
\ende
Our model convection zone is a spherical layer with adiabatic stratification that is heated from below and cooled at the surface. 
We assume an ideal gas with the equations of state,
\begin{equation}
 p= \frac{\cal R}{\mu} \rho T = C_p \frac{\gamma-1}{\gamma} \rho T
\end{equation}
($p$ is the gas pressure, $\cal R$ the gas constant, $C_p$ the heat capacity for constant pressure) 
and hydrostatic equilibrium,
\begin{equation}
 \frac{d p}{d r} = -g \rho.
 \end{equation} 
From the law of gravity,
\begin{equation}
  g = - \frac{GM(r)}{r^2},
 \end{equation}
($G$  gravity constant, $M(r)$ the mass contained in the sphere of radius $r$),
we derive
\begin{equation} \label{simpgrav}
   \frac{{\rm d}g(r)}{{\rm d}r}=-2\frac{g(r)}{r}+4 \pi G \rho.
\end{equation}
For adiabatic stratification,
$
 {p}{\rho^{-\gamma}} = {\rm const.},
 $
the density can be expressed in terms of the temperature,
\begin{equation} \label{simpdens}
  \rho=\rho_{\rm e} \left(\frac{T}{T_{\rm e}}\right)^{1/(\gamma-1)},
\end{equation} 
where $\rho_{\rm e}$ and $T_{\rm e}$ are the values of density and temperature at a reference radius $r_{\rm e}$.
The condition of hydrostatic equilibrium
can then be rewritten in terms of the temperature, i.e.
\begin{equation} \label{simptemp}
  \frac{{\rm d}T}{{\rm d}r}=-\frac{g(r)}{C_p}.
\end{equation}
Equations (\ref{simpgrav}), (\ref{simpdens}), and (\ref{simptemp}) form a system that can be integrated from the reference radius, where the values of density, gravity, and temperature are known from the full stellar evolution model.    
\subsection{Boundary conditions and numerical method}   
As boundary conditions we assume stress-free and impenetrable boundaries for the gas motion and an imposed radial heat flux. At the lower boundary the heat flux is constant: corresponding to the stellar luminosity for the heat transport,
   \begin{equation}
      {F}_r = \frac{L}{4 \pi r_{\rm b}^2}
   \end{equation}
   where $L$ is the stellar luminosity and $r_{\rm b}$ the radius at which the boundary is located. As the radiative heat flux is prescribed and we define the bottom of the convection zone as the point where the radiative heat flux equals the total heat flux this condition is equivalent to imposing zero convective heat flux.
   
At the upper boundary, we assume that the convective heat flux equals the radiative flux is converted into radiation (\cite{gilman81}), which is expressed through the linearized condition 
   \begin{equation}
     {F}_r = \frac{L}{4 \pi r_{\rm t}^2} (1 + 4 \frac{S}{C_p}),
   \end{equation}
where $r_{\rm t}$ is the radius of that boundary.      
The boundary conditions on the rotation axis are implied by the requirements for axisymmetry.
To solve the above system an expansion in terms of spherical harmonics is used for the dependence on latitude.
For the specific entropy, the angular velocity and the stream function $\psi$ of the meridional flow, 
\begin{equation}
	u_r = \frac{ 1}{\rho r^2 \sin \theta} 
	\frac{\partial \psi}{\partial \theta}, \ \ \ \ \ 
	u_\theta = - \frac{1}{\rho r \sin \theta} 
	\frac{\partial \psi}{\partial r } ,
	\label{25}
\end{equation}
we have
\begin{eqnarray}
	S(r,\theta ) &=& \sum_{n=1}^{N}S_n(r)\overline{P}_{2(n-1)}(\cos \theta ),
	\nonumber \\
	\Omega(r,\theta ) &=& \sum_{n=1}^{N} \Omega_n(r)
	\frac{\overline{P}^1_{2n-1}(\cos \theta )}{\sin\theta},
	 \label{26}\\
	\psi (r,\theta ) &=& \sum_{n=1}^N \psi_n(r) 
	\overline{P}^1_{2n} (\cos \theta ) \sin\theta. \nonumber
\end{eqnarray}
In this equation, $\overline{P}_l$ and $\overline{P}^1_l$ are the normalized standard and adjoint Legendre polynomials.
The expansions  are convenient because the functions of colatitude $\theta$ on the right hand side are the eigenfunctions of the angular parts of the corresponding diffusion operators. In using only Legendre polynomials of only even or  odd degree in a certain expansion we assume symmetry with respect to the equator. {The expansion (\ref{26}) usually converges fast such that $N=5$ is sufficient for cases like the solar rotation. Faster rotating stars require values up to 20.}

Applying the expansions (\ref{26}) to the system of partial differential equations (\ref{heat}), (\ref{omega}), and (\ref{curl}) transforms the equations into a system of ordinary nonlinear differential equations with the independent variable $r$. We further reduce the equations to the system of first-order differential equations by introducing new dependent variables. In particular, the Reynolds stress components $R_{r\theta}$ and $R_{r\phi}$ are convenient new dependent variables. The resulting system of first-order differential equations is solved by the standard relaxation method as described by Press et al. (1992). 

The rapidly rotating stars have very thin boundary layers near the top and bottom of their convection zones. To resolve the layers, nonuniform numerical grid (zeros of Chebyshev polynomials) is applied
\begin{eqnarray}
    r_i = \frac{1}{2}\left( r_\mathrm{t} + r_\mathrm{b} -
    \left( r_\mathrm{t}-r_\mathrm{b}\right)
    \cos\left(\pi\frac{i-3/2}{n-2}\right)\right)&,&
    \\
    2 \leq i \leq n-1,\ \ \ \ \ \ \
    r_1 = r_\mathrm{b},\ \ r_n = r_\mathrm{t} &,&
    \nonumber
    \label{28}
\end{eqnarray}
where $n$ is the number of radial grid points. The grid has fine spacing near the boundaries. { To solve for the solar rotation law, a value of $n$ as small as 20 is sufficient, but we usually use higher resolutions (e.g.~$n=200$), especially for fast rotation.} 
\section{The Sun}
%
We first simulate the solar rotation law  and compare the results to those from earlier models by Kitchatinov \& R\"udiger (1999) and K\"uker \& Stix (2001) solving  the same set of  equations with different numerical methods. KR99 used a simple Kramers' law for the opacity while KS01 used the opacity from Ahrens et al.~(1992). Now an analytical opacity law after \cite{stellingwerf75} as distributed with \cite{hansen94} is used. A more elaborate treatment is possible but not likely to improve the model as we do not treat the uppermost layers of the convection zone and the atmosphere.
As the solar convection zone is not fully ionized up to the photosphere, the adiabatic gradient,
 $
  \nabla_{\rm ad} = \left({\partial \log T}/{\partial \log P}\right)_{S},
 $
and hence the specific heat capacity are not constant. 
While $\nabla_{\rm ad}=0.4$ holds throughout most of the convection zone, it drops to values as low as 0.1 in the upper 35,000 km. As we work with constant values we either have to exclude the uppermost layer or adjust $\nabla_{\rm ad}$ to reproduce the depth of the convection zone from the stellar evolution model.
\begin{figure}[htb]
\includegraphics[width=9.0cm]{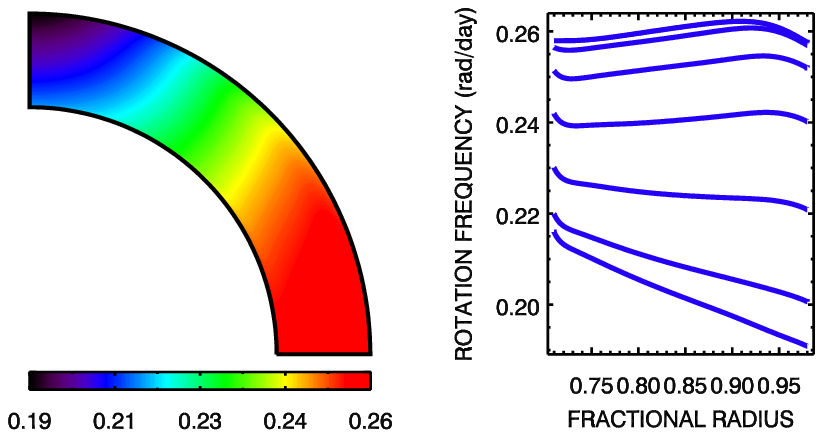}
\includegraphics[width=9.0cm]{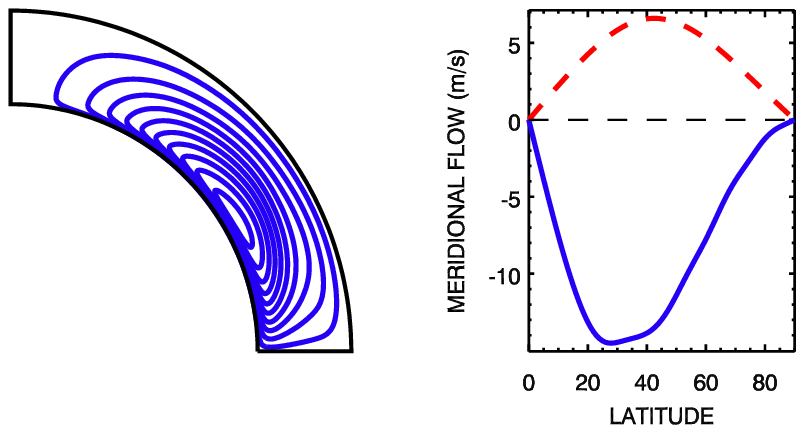}
\caption{ \label{solar}
The solar differential rotation and meridional flow as computed with the new scheme. Top left: Surface rotation rate vs. co-latitude.  Top right: rotation rate as function of radius at 0, 15, 30, 45, 60, 75, and 90$^\circ$ latitude, respectively, from top to bottom.
Bottom left: Streamlines of the meridional flow. The flow is counter-clockwise, i.e. directed towards the pole at the surface and towards the equator at the bottom.  Bottom right: Flow speed as function of latitude at the top (solid blue) and bottom (dashed red) of the convection zone. Positive values indicate a flow away from the north pole.
}
\end{figure}
%
%
%
\begin{figure}[htb]
\begin{center}
  \includegraphics[width=8.0cm]{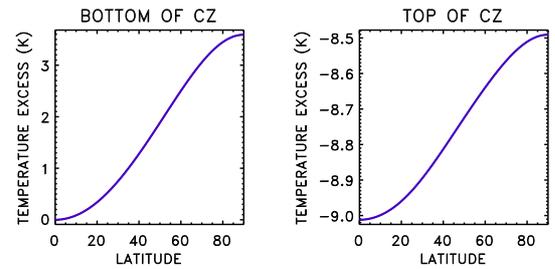}
  \end{center}
  \caption{ \label{suntemp}
  Temperature deviation as function of the latitude at the bottom (left) and top (right) of the solar convection zone. The polar axis is always warmer than the equator. The differences, however, are small.
   }
\end{figure}

The top panels in Fig.~\ref{solar} show the rotation pattern in the solar convection zone. It agrees well with those computed with the KR99 codes and the KS01 scheme with the modifications described in Bonanno et al.~(2007). The surface rotation is fastest at the equator with a difference 
\bege
 \delta \Omega = 0.07\ {\rm rad/day}
\ende
between the rotation rates of the equator and the polar caps, corresponding to a relative shear, $\delta \Omega / \Omega_{\rm eq},$ of 29 percent. The variation with radius at midlatitudes is weak. The angular velocity decreases with radius at high latitudes while at the equator it slightly increases with radius (`superrotation') in the deep convection zone. In the surface layer there is a  negative gradient of $\Omega$ (`subrotation') at all latitudes. Like the rotation profiles computed with the KR99 and KS01 schemes, this rotation profile is in excellent agreement with the findings of helioseismology. 
 
The bottom panels of Fig. \ref{solar} show the meridional flow. There is one  cell per hemisphere with the surface flow directed towards the poles and the return flow at the bottom of the convection zone. The  amplitude of this `counter-clockwise' flow is 14 m/s at the (model) surface and 6 m/s at the bottom. The difference between the flow speeds at the top and bottom respectively is smaller than might be expected from the density stratification and the requirement of mass conservation but the concentration of the return flow to a thin layer allows a relatively fast flow despite the larger mass density. 

{
Figure \ref{suntemp} shows the quantity 
\bege
  \delta T = (S-S_0)\frac{T}{C_p},
\ende
at the top and the bottom of the solar convection zone with $S_0$  the specific entropy at the bottom of the convection zone at the equator. This choice of $S0$ implies negative values of $\delta T$ at the top of the convection zone. In general, the polar axis is warmer than the more equatorward parts of the fluid. Unfortunately, the smallness of the temperature difference does not allow empirical confirmations. Though the deviation from the adiabatic stratification is only small, the resulting barocline term  has a profound impact on the large-scale meridional flow and the differential rotation (see below). 
}
\subsection{Fast-rotating Sun}
\begin{figure}[htb]
\begin{center}
\includegraphics[width=8.0cm,height=4.0cm,]{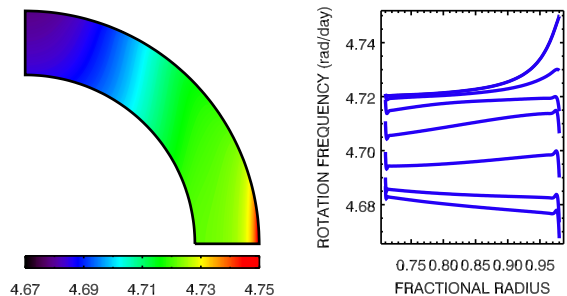}
\end{center}
\caption{\label{fastsun}
Rotation of a hypothetical fast-rotating Sun with $P= 1.33$ d.}
\end{figure}

To study the impact of the basic rotation we compute the  rotation law of a hypothetical fast-rotating Sun with the $P=1.33$ day rotation period of HD 171488. Figure \ref{fastsun} shows the resulting rotation pattern. The isocontours are cylinder-shaped in accord to the Taylor-Proudman theorem. Consequently, the radial profiles in the right diagram show an increase of the rotation rate with increasing radius at low latitudes. Unlike the case of the real Sun, there is a pronounced increase in the upper part of the convection zone at the equator. At both radial boundaries there are pronounced boundary layers with huge gradients of the rotation rate which are caused by the stress-free boundary conditions. The surface rotation is fastest at the equator and decreases monotonously towards the polar caps but the slope of the decline has a minimum at mid-latitudes. With a value of 0.08 rad/day instead of 0.07 rad/day the horizontal shear is only slightly stronger than for the Sun despite a factor of 20 between the average rotation rates. 

The meridional flow only has one flow cell per hemisphere with the surface flow directed towards the poles. The amplitudes are 34 m/s at the surface and 17 m/s at the bottom of the convection zone. As for the real Sun with its much slower rotation, the return flow is half as fast as the surface flow but even more concentrated at the bottom of the convection zone. The flow is fastest at mid-latitudes. 

%
%
%
\subsection{MOST stars:  $\epsilon$ Eri and $\kappa^1$ Cet }
$\epsilon$ Eri and $\kappa^1$ Cet are active dwarf stars for which surface differential rotation has been derived from light curves recorded  by the MOST satellite 
Croll et al.~(2006) derived a rotation law of the form
\bege
  P(B) = \frac{P_{\rm eq}}{1-k\sin^2 B} 
\ende
for the K2 V star $\epsilon$ Eri,
where $P(B)$ is the rotation period at the latitude $B$, and $P_{\rm eq}$ the rotation period at the equator ($B=0$). The parameter $k$  gives the shear  of the  rotation law. The general rule derived from the theory of the stellar rotation laws  of Kitchatinov \& R\"udiger (1999) 
\bege
k \simeq \frac{P_{\rm eq}}{100\ \rm days}
\label{pred}
\ende
is perfectly fulfilled by these stars.

%
An equatorial rotation period $P_{\rm eq} = 11.2$ days and a value $k=0.11^{+0.03}_{-0.02}$ for the differential rotation has been reported for  $\epsilon$ Eri corresponding  with $\delta \Omega = 0.062$ rad/day.   For the G5 V star $\kappa^1$ Cet
Walker et al.~(2007) found $P_{\rm eq} =8.77$ days and $k = 0.09^{+0.006}_{-0.005}$ corresponding with $\delta \Omega=0.064$ rad/day.

Kitchatinov \& Olemskoy (2010) computed rotation laws for model stars with 0.8 and 1.0 solar masses for $\epsilon$ Eri and $\kappa^1$ Cet. The results were  $k=0.127$ for $\epsilon$ Eri and $k=0.13$ for $\kappa^1$ Cet. 

Here the  calculations for these stars are based on improved stellar models computed with the MESA/STARS code (\cite{paxton}). 
A star with $M_\star=0.85 M_\odot$, $Z=0.02$ and an age of 0.44 Gyr serves as model for $\epsilon$ Eri. It has an effective temperature of 5076 K, a radius of 0.76 $R_\odot$, and a luminosity of 0.34 $L_\odot$. The bottom of the outer convection zone is located at 0.69 $R_\star$. For this model star we find a surface differential rotation $\delta \Omega=0.51$ rad/day or $k=0.10$. This is in agreement with the observed value.

For $\kappa^1$ Cet we use solar mass and metalicity and an age of 1 Gyr. The model star has an effective temperature of 5677 K,  0.915 $R_\odot$ and    0.78 $L_\odot$. The resulting rotation law shows a surface shear $\delta \Omega = 0.077$ rad/day, or $k=0.12$. This is  slightly  larger than the observed value. An earlier model from the same evolutionary track, with a stellar age of 167 Myr, an effective temperature of 5649 K, a radius of 0.9 $R_\odot$, and an luminosity of 0.74 $L_\odot$ yields $k=0.11$, or $\delta \Omega=0.072$ rad/day. This is still not  in perfect agreement with the observed value but reasonably close.

Possible reasons for the remaining discrepancy are the stellar model, which might not be a sufficiently accurate representation of the real star and an underestimation of the total surface shear by the $\sin^2\theta$ derived from spot rotation as the observed spots  do not cover the whole range of latitudes from pole to equator.
%
\section{Numerical experiments}
We  carry out several numerical experiments to compare the roles played by the two drivers of differential rotation, i.e. the $\Lambda$ effect and the barocline flow.
\subsection{Sun without $\Lambda$ effect}
Our model includes two effects that are capable to maintain differential rotation, the $\Lambda$ effect and the barocline flow as due to a horizontal temperature gradient. This can be seen when the barocline term in Eq. (\ref{curl}) is rewritten in terms of the temperature,
\bege \label{baro}
   \frac{1}{\rho^2}(\nabla \rho \times \nabla p)_\phi 
     \approx - \frac{g}{rT} \frac{\partial \delta T}{\partial \theta}.
\ende
The barocline term can be a powerful driver of meridional flows, which in turn can drive differential rotation. In our theory the latitudinal temperature profile is due to the anisotropic heat-conductivity tensor in the presence of rotation. Always the polar axis is slightly warmer than the equator but with an unobservable temperature excess.
A positive pole-equator temperature difference will drive a {\em clockwise} flow. Its angular momentum transport  leads to an accelerated rotation at low latitudes and to slower rotation at the polar caps. A barocline flow (also  `thermal wind') can therefore maintain  differential rotation with solar-type surface rotation.
For an illustration  we repeat our computation for the Sun  with the $\Lambda$  terms canceled within the Reynolds stress. 

\begin{figure}[htb]
\begin{center}
  \includegraphics[width=8cm,height=4cm]{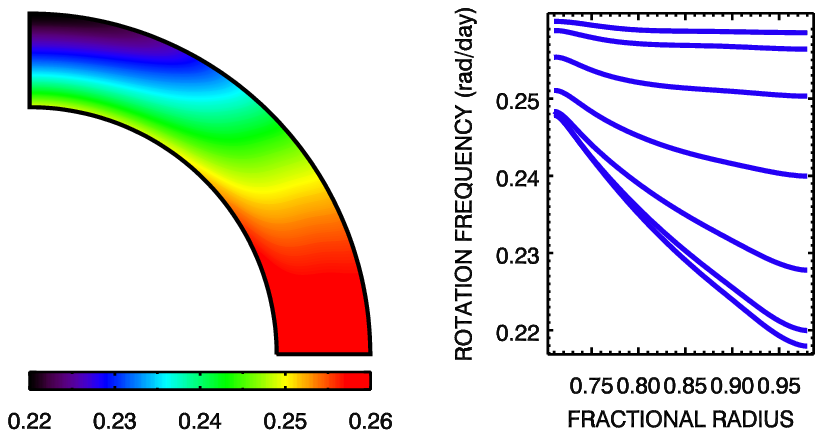}  
  \includegraphics[width=8cm,height=4cm]{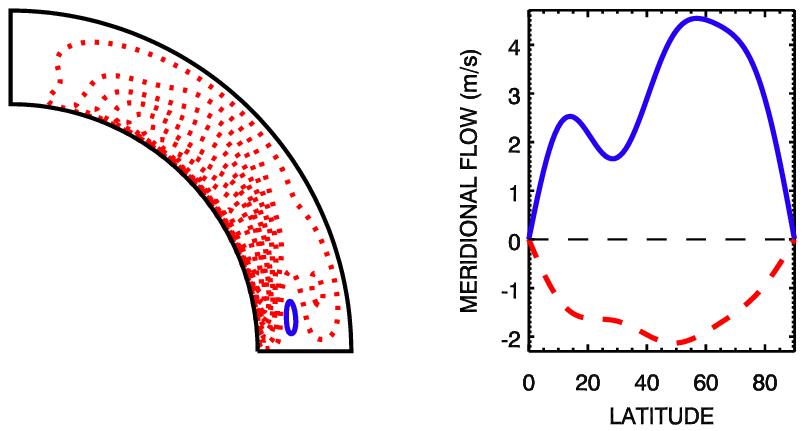} 
   \end{center}
  \caption{ \label{solnolamdr}
   Solar rotation law and meridional flow without  $\Lambda$ effect. Top: The rotation law. Note the steep negative gradient of the angular velocity at the poles and the almost rigid rotation beneath the equator. Bottom: 
   The meridional flow   is   anti-solar: it is positive (equatorwards) at the top of the convection zone and negative (polwards) at the bottom of the convection zone.}
\end{figure}

The resulting rotation and flow patterns are shown in Fig. \ref{solnolamdr}. The rotation is indeed solar-type but with $\delta \Omega = 0.04$  weaker than with $\Lambda$ effect and the isocontours are distinctly disc-shaped at the poles even in the midlatitudes. The equator region shows basically no structure. 

The typical superrotation of the deep convection zone beneath the equator  known as a result of the helioseismology does not occur. The reason is simple: if only a (fast) meridional circulation transports  the angular momentum with a  pattern  symmetric with respect to the equator then the angular momentum becomes uniform in the equatorial part of the convection zone  (except, of course,   the boundary layers). Hence, there is $r^2\Omega\simeq \rm const$ independent of the flow direction but in contradiction to the observation \footnote{without $\Lambda$ effect a  superrotation beneath the equator can only be due to an  anticlockwise flow which is sufficiently slow}.

With amplitudes of 4.7 m/s at the top and 2.1 m/s at the bottom of the convection zone the barocline-induced clockwise meridional flow is substantially weaker than in the complete  model -- and it goes  in the `wrong' direction.
%
\subsection{Sun without barocline flow}
\begin{figure}[htb]
\begin{center}
  \includegraphics[width=8cm,height=4cm]{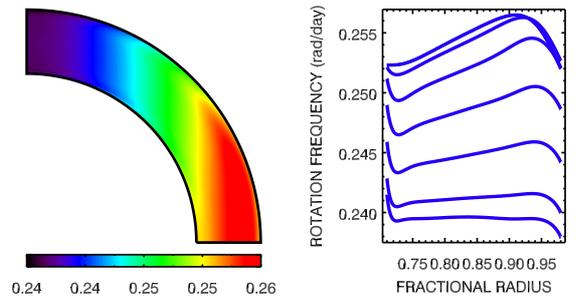} 
\end{center}
  \caption{ \label{solnobarodr}
  Solar rotation law without  baroclinic terms.  The structure disappears now at the polar axis but it  appears at the equatorial region. The equatorial plane shows superrotation but the polar axis rotates almost rigid.}
\end{figure}

{
Next  the barocline term is canceled while keeping the $\Lambda$ effect which  maintains  the differential rotation together with the meridional flow caused by the former through the second term on the LHS of Eq.~(\ref{curl}) (`Biermann-Kippenhahn flow', see Kippenhahn 1963; K\"ohler 1970). 
}
Figure \ref{solnobarodr} shows the resulting rotation pattern. The differential rotation is solar-type, i.e. the rotation period is shortest at the equator and longest at the polar caps. With $\delta \Omega = 0.014$ rad/day the surface rotation is much more rigid than observed. The isocontour plot shows a distinctly cylinder-shaped pattern in the bulk of the convection zone while at the top and bottom boundaries the pattern deviates from the cylinder geometry and the rotation rate falls off with increasing radius at all latitudes. The {\em counterclockwise} meridional flow has a very similar geometry   as in the full model but it is faster with amplitudes of 17 and 8 m/s at the top and bottom.  

The subrotation along the polar axis which is typical for the solar rotation law and which is  produced by baroclinic flows (cf. Fig. \ref{solnolamdr}) does not exist here.  K\"ohler (1970) and  R\"udiger et al. (1998) have given several numerical   examples of this direct consequence of the Taylor-Proudman theorem.

\section{Young G dwarfs}
\subsection{CoRoT-2a}
This G7 V star is a young Sun, i.e. it has solar mass but is much younger with an age of 0.5 Gyrs. The star has been observed by the CoRoT satellite and found to have a planet with an orbital period of 1.743 days (Alonso et al.~2008). Besides planetary transits, the light curve shows periodic variation that is most easily explained by a rotating, spotted surface with basic rotation period of 4.5 days. Spot modeling using circular spots finds an excellent fit assuming three spots and a solar-type surface differential rotation with a difference of 0.11 rad/day (Fr\"ohlich et al.~2009). 
\begin{figure}[htb]
\begin{center}
\includegraphics[width=8.0cm,height=4.0cm]{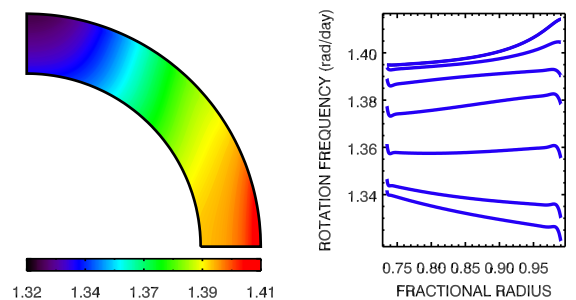}
\end{center}
\caption{\label{corot2a}
Rotation pattern of CoRoT-2a, a young Sun rotating with  $P=4.5$ days.}
\end{figure}

 Our model star is based on a model from an evolutionary track for the 
Sun. The age is 0.5 Gyrs, the radius 0.9 solar radii, and the luminosity 75 percent of the solar value. The bottom of the convection zone is at a fractional radius $x=0.73$.

Figure \ref{corot2a} shows the resulting rotation pattern. The surface differential rotation of 0.09 rad/day  reproduces the value observed by   Fr\"ohlich et al.~(2009). The rotation pattern is more cylindrical than that of the Sun, but less than that of our fast-rotating Sun. Similarly, the boundary layers are more pronounced than in the Sun but less  than for the fast-rotating Sun. The flow pattern is similar to that in the solar convection zone. The amplitude is slightly larger with a maximum value of 18.6 m/s at the top and 10.6 m/s at the bottom of the convection zone. 

\subsection{ R58, LQ Lup  and HD 171488}
We now address the recent differential rotation measurements of the fast-rotating G dwarfs  R58, LQ Lup and HD 171488. 
R58 (HD 307938) is a young active G dwarf in the open cluster IC 2602. It has a photospheric temperature of 5800 K and rotates with a period of 0.56 days (\cite{marsden05}).
LQ Lup (RX J1508.6-4423) is a post-T Tauri star with an effective temperature of $5750\pm50$ K and a rotation period of 0.31 days, and a radius of $1.22\pm0.12$ solar radii (\cite{donati2000}), corresponding to a mass of $1.16\pm0.04$ solar masses, and an age of $25\pm10$ Myr. The equator-pole $\Omega$-difference is 0.12  rad/day.
HD 171488 (V889 Her) is a young active G dwarf. Strassmeier et al.~(2003) found a photospheric temperature of 5830 K, a rotation period of 1.337 days, a mass of $1.06\pm0.05$ solar masses, a radius of $1.09\pm0.05$ solar radii, and an age of 30--50 Myr.  Marsden et al.~(2006) find a rotation period of 1.313 days, a photospheric temperature of 5800K, and a radius of $1.15\pm0.08$ solar radii.

With 0.4--0.5 rad/day the equator-pole $\Omega$-difference of HD171488 is exceptionally large. The $k$-value of 0.10 misses the $k=0.013$ predicted by Eq. (\ref{pred}) by a factor of 7.5. As the corresponding factors for R58 and LQ Lup are much smaller (factor 2--3)  the following calculations take the case of HD 171488 as the main example.
As the stars have similar radii and effective temperatures we investigate what impact differences in the internal structure have on the surface rotation patterns. We study two models that mainly differ in the metalicity and, as a consequence, the depth of the convection zone. The models were computed with the MESA/STARS code. We use the metalicity as control parameter for the depth of the convection zone and vary mass and mixing length parameter to adjust radius and effective temperature to values close to those observed for the three young G dwarfs. The rotation period is the 1.33 day period of HD 171488. Both model stars are 30 Myr old.
%
\subsubsection{Deep convection zone}
\begin{figure}[htb]
\begin{center}
\includegraphics[width=8.0cm,height=4.0cm]{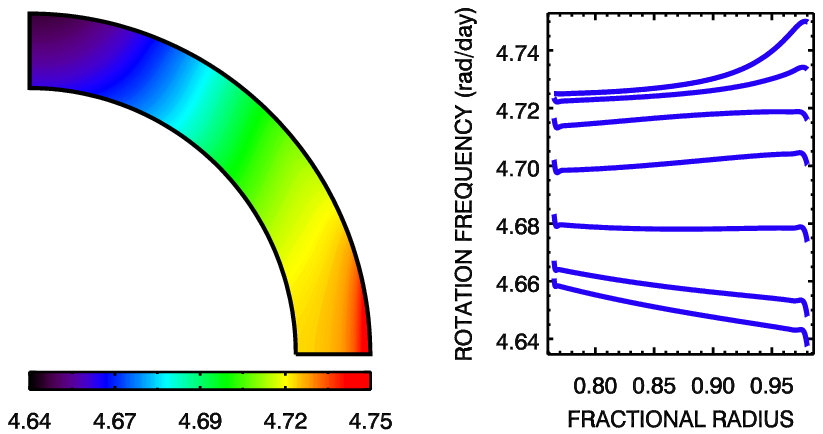}
\includegraphics[width=8.0cm,height=4.0cm]{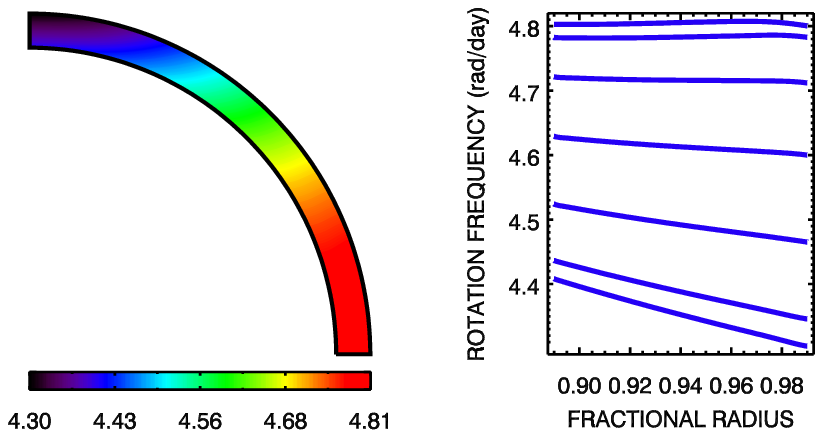}
\end{center}
\caption{\label{deepdr}
Internal rotation of young G dwarfs with rotation period of 1.33 days. Top: deep convection zone, $\delta\Omega=0.11$ rad/day at the surface . Bottom: shallow convection zone. $\delta\Omega=0.5$ rad/day at the surface.  
}
\end{figure}

The  first model star has the high metalicity of $Z=0.03$. A mass of 1.11 solar masses and a value of 1.6 for the mixing length parameter lead to an effective temperature of 5685 K and a radius of 1.14 solar radii. The bottom of the convection zone is located at $x=0.765$ and has a temperature of $1.46\times 10^6$ K.
The rotation law is shown in the top part of Fig.~\ref{deepdr}. The isocontour plot shows cylinder-shaped contours similar to those for the fast-rotating Sun. The surface rotation law is solar-type with a fast-rotating equator. The surface shear is much stronger. We find a value of 0.11 rad/day, about 1.5 times the solar value. The radial profiles show superrotation beneath the equator and subrotation along the rotation axis.  The boundary layers are much less pronounced and the surface layer resembles the real Sun, i.e.~the rotation rate decreases with increasing radius at all latitudes.
The meridional flow is of solar-type (counterclockwise) with the surface flow towards the poles and the return flow located at the bottom of the convection zone. The amplitudes are 23 and 14 m/s, respectively. 

\subsubsection{Shallow convection zone}
Our second model star has a metalicity of 0.01. A mass of 1.08 solar masses and a mixing length parameter of 1.0 lead to a radius of 1.14 solar radii and an effective temperature of 5750 K. The bottom of the convection zone lies at a fractional radius of 0.89 and a temperature of $6\times10^5$ K.
The bottom part of Fig.~\ref{deepdr} shows the resulting rotation law. The surface rotation has a total shear of $\delta \Omega = 0.5$ rad/day. The contour plot shows a disc-shaped pattern with flat profiles at low latitudes and a decrease of the rotation rate with increasing radius at the polar caps. 
The radial profiles show pronounced boundary layers at the radial boundaries.
The flow pattern is mostly solar-type. There is one large flow cell with poleward flow at the top and the return flow at the bottom of the convection zone. In addition, there is a small cell of clockwise flow at low latitudes. The flow speeds at the top reach 18 m/s and 2.2 m/s, respectively. The return flow at the bottom of the convection zone reaches 3.5 m/s. 
%
%
%
%

%
%
%
\subsubsection{Truncated convection zone}
%
%
The models above have (roughly) the same effective temperature but differ in mass and structure. 
To further isolate the effect of the depth of the convection zone, we compute a model of a young solar mass star and artificially reduce the depth of the convection zone by imposing the lower boundary at a fractional radius $x=0.88$ instead of $x=0.78$ and compare the rotation pattern with that of the full model. In both cases the top boundary is at $x=0.98$. The truncated convection zone thus has half the depth of the full model.
In both cases the rotation law is solar-type but while the full convection zone produces a surface shear of 0.12 rad/day, the truncated convection zone has 0.32 rad/day. The full model has a temperature difference of 24 K at the top and 124 K at the bottom. In case of the truncated model the polar cap is 47 K hotter than the equator at the top and 254 K at the bottom. 
\subsubsection{Very fast rotation}
\begin{figure}[htb]
\begin{center}
   \includegraphics[width=8cm,height=4cm]{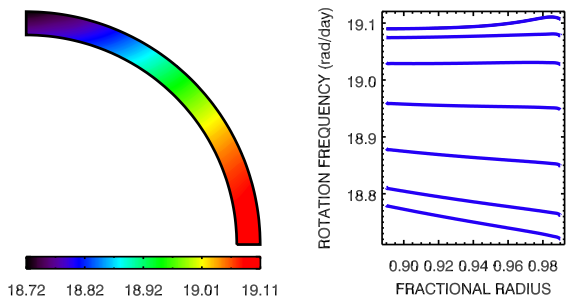}
  \end{center}
  \caption{ \label{shallowfastrot}
    Differential rotation of the G dwarf with the shallow convection zone and a rotation period of 0.33 days. Left: isocontours of the angular velocity. Right: the rotation profiles at various latitudes.
    }
\end{figure}
To illustrate the impact of the rotation rate, we repeat the computation for the shallow convection zone model with rotation period of 0.33 days. Figure \ref{shallowfastrot} shows the resulting rotation pattern. The left plot shows solar-type rotation  both in the surface differential rotation and in the predominantly radial isocontours. This is still not cylinder-shaped but closer to the Taylor-Proudman than the disc-shaped pattern we found for a rotation period of 1.33 days. The surface shear is 0.38 rad/day, the temperature difference between polar caps and equator is 1785 K at the bottom and 184 K at the top. The maximum flow speeds are 14 m/s at the top and 8 m/s at the bottom of the convection zone.

%
%
%
%
%
%
\section{Discussion}
\subsection{Shallow vs.~deep convection zone}
Our ``shallow" and ``deep" convection zone models have similar radii and effective temperatures but the stars differ in mass and metalicity. We have also chosen different values for the mixing-length parameter. As a result, the depth of the convection zone differs between the models. We find very different results for the differential rotation of these convection zones. The model with a shallow convection zone shows much stronger differential rotation  than the ones with deeper convection zones. Our ``shallow convection zone" model does indeed reproduce the very large value of 0.5 rad/day found by Jeffers \& Donati (2009) while the ``deep convection zone" model with its smaller value of 0.11 is in rough agreement with the values observed for LQ Lup and R58. 

The result from the shallow convection zone  model is in line with previous findings on F stars. Extreme values of surface shear have been observed for F stars by Reiners (2006) and found in theoretical models for stars with very shallow convection zones by K\"uker \& R\"udiger (2007). In the latter study, the largest values of surface shear are found for the most massive stars studied. These stars are not only the hottest, their convection zones are also very shallow. 

To achieve the thin convection zone we had to lower the mixing-length parameter to a value of 1.0. This is not only different from the value used for the ``deep" convection zone model but much lower than what is usually used in stellar evolution models. Hence, the model is probably not realistic and does not directly apply to LQ Lup. Not being specialists in the field, we leave the question whether a shallow convection zone is possible open but we wonder if strong magnetic fields could inhibit the convective heat transport in a way that might be mimicked by this choice of parameter.

The ``truncated convection zone" model avoids the uncertainties about the "shallow convection zone" model and shows even more clearly the impact the depth of the convection zone has on the surface rotation.  
Both the vertical and horizontal temperature gradients are steepest for the model with the shallow convection zone and flattest for the model with the deep convection zone. At the bottom of the convective zone the differences are 750 K and 100 K for a convection zone of small or large depth, respectively. The corresponding values at the top are 55 K and 19 K. 
\subsection{Thermal wind equilibrium}

We always find a higher temperature at high latitudes than at the equator. This is a consequence of the tilt of the convective heat transport vector towards the axis of rotation caused by the Coriolis force. 
In spherical polar coordinates the components of the heat flux read
\begin{eqnarray} 
   F_r &=& -\rho T \chi_{r r} \frac{\partial S}{\partial r} - \rho T \chi_{r \theta} \frac{\partial S}{\partial \theta}, \\
    F_\theta &=& - \rho T \chi_{\theta r} \frac{\partial S}{\partial r} - \rho T \chi_{\theta \theta} \frac{\partial S}{\partial \theta}.
\end{eqnarray}
The first term in the horizontal component precludes a purely radial stratification. Any variation of the specific entropy with the radius will cause a horizontal heat flux and thus build up a horizontal gradient. 
This case is profoundly different from that of a latitude-dependent but still purely radial heat flux. The latter would have a much smaller impact and would result in a much weaker differential rotation (\cite{rue05}). 

The impact on the maintenance of differential rotation can be seen from the equation for the meridional flow. For fast rotation (i.e. large Taylor number) Reynolds stress and nonlinear terms can be neglected and Eq. (\ref{curl}) is reduced to the thermal wind equation
\bege \label{fastrot}
   2 r \sin \theta \Omega \frac{\partial \Omega}{\partial z} - \frac{g}{r T} \frac{\partial \delta T}{\partial \theta} = 0.
\ende 
 It follows that a gradient of the angular velocity along the axis of rotation is needed to balance the horizontal temperature gradient. Hence, a disc-shaped rotation pattern results in the polar area (only) due to the action of the baroclinic flow. For warmer poles ($\delta T$ sinks equatorwards) the $z$-gradient of $\Omega$ is negative as observed. The meridional flow  without the baroclinic component only yields ${\rm d}\Omega/{\rm d}z \simeq 0$ (K\"ohler 1970, see also Fig. \ref{solnobarodr}). Hence, the empirical finding  of 
 negative ${\rm d}\Omega/{\rm d}z$ at the polar axis by helioseismology proves the existence of baroclinic flows  in the solar convection zone.

As we have demonstrated by means of Fig. \ref{solnolamdr} the superrotation beneath the solar equator is a direct indication for the action  of the Reynolds stress  in the convection zone. All models without $\Lambda$-effect but with an clockwise (equatorward at the surface) circulation lead to ${\rm d}\Omega/{\rm d} r \lsim 0$ in the bulk of the convection zone beneath close to the equator. Such clockwise flows are able to accelerate the equator compared with the poles but they cannot produce the superrotation beneath the equator.

\begin{figure}[htb]
\begin{center}
  \includegraphics[width=4.4cm]{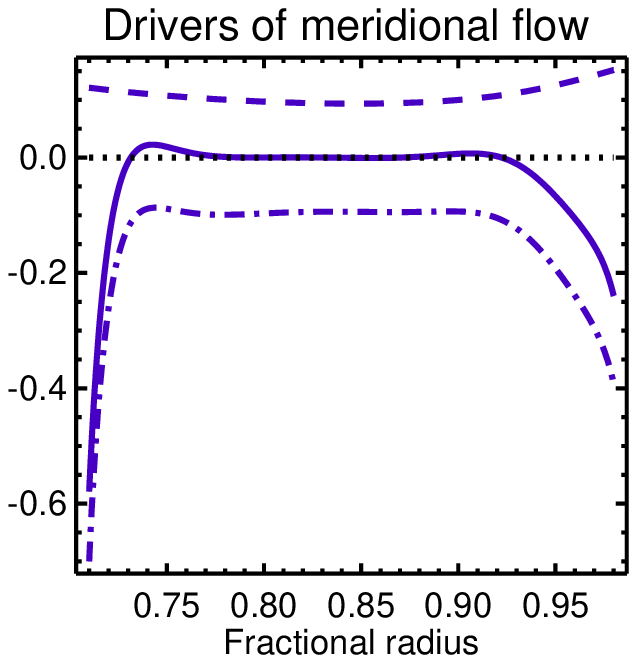}
  \includegraphics[width=4.4cm]{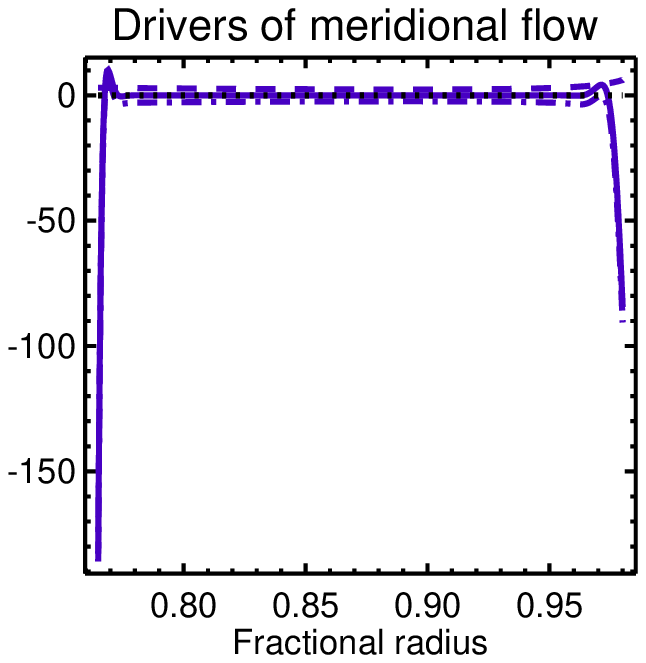}
\end{center}
\caption{ \label{suntherm}
  The drivers of angular momentum vs. fractional radius at 45$^\circ$  latitude for the Sun (left, slow rotation) and the deep convection zone G dwarf (right, fast rotation). Dashed line: baroclinic term. Dash-dotted line: centrifugal term. Solid line: both. Note that the balance (\ref{fastrot}) well holds in the bulk of the convection zones of fast rotation stars where the meridional circulation is slow and smooth. Equation (\ref{fastrot}) does not hold in the boundary areas where the flow is fast and shearing.
  }
 \end{figure}

Figure \ref{suntherm} shows results from a detailed computation of the terms on the LHS of Eq.~(\ref{fastrot}) and their sum vs.~the fractional radius for a latitude of 45$^\circ$. The left panel shows the plot for the Sun, the right panel that for the deep convection zone G dwarf model. The same quantities were computed for the fast Sun and the shallow convection zone G dwarf models.

For the real Sun the two terms cancel in the bulk of the solar convection zone where the meridional flow is slow and smooth but not close to the boundaries. For the deep convection zone the boundary layers are much thinner but also much more pronounced. 
These findings indicate that for fast rotation the meridional flow is mainly driven by the boundary layers while the bulk of the convection zone is in thermal wind equilibrium.  In the boundary layers the meridional circulation is strongly sheared so that the RHS of Eq.~(\ref{fastrot}) (which consists on $3^{\rm rd}$ derivatives of the meridional velocity components  multiplied with the eddy viscosity) becomes large.

 %
 %
The numerical experiments show that the baroclinic term is an important but not the sole driver of the solar differential rotation. Cancellation of the $\Lambda$ effect leaves a surface shear of 0.04 rad/day while switching off the baroclinic term reduces it to 0.014 rad/day. This is not because the $\Lambda$ effect is an inefficient transporter of angular momentum but because it is balanced by the meridional flow. Canceling  the meridional flow altogether increases the surface shear to 0.18 rad/day. This is a well-known effect: meridional flows driven by differential rotation act to reduce it (\cite{rue98}). The baroclinic flow has the opposite direction and builds up the differential rotation along the polar axis. The observed superrotation beneath the equator, however,  cannot be produced by meridional circulation but should be a direct consequence of the existence of the $\Lambda$ effect.
%
%
%
%
\section{Conclusions}
Stellar differential rotation is driven by Reynolds stress, by the centrifugal-induced (`Biermann-Kippenhahn') flow  and (for stratified density) by the  baroclinic flow. In our model both meridional circulations have opposite directions. The barocline flow  becomes important  for faster rotation as the convective heat-flux deviates from the radial direction. This deviation can be interpreted as a tilt of the heat-flux vector towards the rotation axis and causes  the poles to be slightly warmer  than the  lower latitudes. The baroclinic flow is responsible for the strong negative gradient ${\rm d}\Omega/{\rm d} z$ along the rotation axis, and the Reynolds stress produces the typical positive ${\rm d}\Omega/{\rm d} r$ along the equatorial midplane both known from   helioseismology. As a result of both impacts the isolines of $\Omega$ known for the Sun are almost radial in mid-latitudes.

Studying real stellar models implies varying a bunch of parameters  as mass, radius, effective temperature and the depth of the convection zone are interdependent.   
Kitchatinov \& Olemskoy (2010) studied differential rotation along the lower ZAMS for fixed rotation rate and found that the surface shear is a function of the effective temperature alone. Using some  artificial models, we find that for the same effective temperature the depth of the convection zone has a big impact on stellar rotation. Shallow convection zones produce stronger surface shear. 

The presented mean-field theory for G stars naturally  explains the rotation laws of the Sun, of the MOST stars $\epsilon$ Eri and $\kappa^1$ Cet, and of such fast-rotating stars like  R58 and LQ Lup.
The discrepancy between these stars and the  much stronger differential rotation observed for HD 171488 -- if real -- is hard to explain for stars of similar age and spectral type. If, however, HD 171488 had a shallower convection zone for some reason, its strong surface differential rotation follows  immediately. 

During the pre-MS evolution the convection zone retreats from the central region and the originally fully convective star forms a radiative zone around the core. This change in the depth of the convection zone should be reflected by the surface differential rotation. 

%
%
\begin{acknowledgements}
This work has been funded by the Deutsche Forschungsgemeinschaft (RU 488/21). LLK is thankful to the support by the Russian Foundation for
Basic Research (projects 10-02-00148, 10-02-00391).
\end{acknowledgements}

%
%
%

\end{document}